\documentclass[usegraphicx,useAMS,usenatbib]{mn2e}

\usepackage{amsmath}
\usepackage{aas_macros}

\usepackage[svgnames]{xcolor}

\usepackage{xspace}

\usepackage{hyperref}

\newcommand{\cm}{\textrm{cm}}

\newcommand{\kms}{\textrm{km s}^{-1}}

\newcommand{\sek}{\textrm{s}}

\newcommand{\grm}{\textrm{g}}

\newcommand{\MeV}{\textrm{MeV}}
\newcommand{\keV}{\textrm{keV}}

\newcommand{\etal}{\it et al.\xspace}
\newcommand{\ie}{i.e.\xspace}

\newcommand{\eg}{e.g.\xspace}

\newcommand{\AZNucleus}[3]{{}^{#1}_{#2}\mathrm{#3}}

\newcommand{\msol}{M_{\odot}}

\newcommand{\figref}[1]{Fig.\,\ref{#1}}
\newcommand{\tabref}[1]{Tab.\,\ref{#1}}

\newcommand{\Ti}{$\AZNucleus{44}{}{Ti}$\xspace}
\newcommand{\Ni}{$\AZNucleus{56}{}{Ni}$\xspace}
\newcommand{\mum}{$\mu \mathrm{m}$\xspace}
\newcommand{\MgSiO}{$\mathrm{Mg}_2 \mathrm{SiO}_4$}
\newcommand{\FeO}{$\mathrm{Fe}_3 \mathrm{O}_4$}

\begin{document}

\title
{Titanium hidden in dust}

\author[Iyudin \etal]{
  A.F.~Iyudin$^{1}$, E.~M{\"u}ller$^{2}$, M.~Obergaulinger$^{3,4}$
  \\
  $^{1}$Skobeltsyn Institute of Nuclear Physics, Moscow State
  University by M.V. Lomonosov
  \\
  $^2$ Max-Planck-Institut f{\"u}r Astrophysik, Karl-Schwarzschild-Str. 1,
  85741 Garching, Germany
  \\
  $^{3}$ Institut f{\"u}r Kernphysik, Technische Universit{\"a}t
  Darmstadt, Schlossgartenstraße 2, 64289 Darmstadt, Germany
  \\
  $^4$ Departament d{\'{}}Astronomia i Astrof{\'i}sica, Universitat de
  Val{\`e}ncia, \\ Edifici d{\'{}}Investigaci{\'o} Jeroni Munyoz, C/
  Dr.~Moliner, 50, E-46100 Burjassot (Val{\`e}ncia), Spain
}

\maketitle

\begin{abstract}
  Cassiopeia A, one of the most intriguing galactic supernova
  remnants, has been a target of many observational efforts including
  most recent observations by ALMA, Hubble, Herschel, Spitzer, NuSTAR,
  Integral, and other observatories.  We use recent gamma-ray lines
  observations of the radioactive products of Cas\,A supernova
  explosive nucleosynthesis as well as spectral energy densities
  derived for Cas\,A at infrared wavelengths to speculate about the
  possibility of radioactive \Ti being locked into large dust
  grains. This suggestion is also supported by the possible
  observation of a pre-supernova outburst about 80 years before the
  actual Cas\,A supernova explosion in 1671 AD by Italian astronomer
  G.D.\,Cassini.  The plausibility of such a scenario is discussed
  also with reference to recent supernovae, and to the contribution of
  core-collapse supernovae to the overall dust production in the
  Galaxy.
\end{abstract}


\section{Introduction}
\label{Sek:Intro}

We consider the connection between the line emission of radioactive
isotopes in supernova (SN) ejecta, and the dust produced before and
after the SN explosion. To the best candidates to probe SN explosion
models belong radioactive isotopes like $\AZNucleus{56}{}{Ni}$,
$\AZNucleus{57}{}{Co}$, and \Ti
\citep{Clayton_et_al__1992__apjl__TheCo-57abundanceinSN1987A,Iyudin_et_al__1999__AstrophysicalLettersandCommunications__COMPTELAll-SkySurveyin44TILineEmission,Diehl_et_al__2011__LNP}.
Among these isotopes, radioactive \Ti has the longest lifetime of
about 86 years
\citep[\eg,][]{Ahmad_et_al__2006__prc__ImprovedmeasurementoftheTi44half-lifefroma14-yearlongstudy}.
It is produced mainly during the so-called $\alpha$-rich freeze-out in
core-collapse SNe (CCSNe) with a yield that depends very sensitively
on the conditions in the central zones of the SN, like the mass-cut,
the kinetic energy, and the asymmetry of the explosion. Some amount of
\Ti is also produced during O-shell burning and Si-shell burning.

Observationally, the important signatures of the \Ti decay chain, \Ti
$\to \AZNucleus{44}{}{Sc} \to \AZNucleus{44}{}{Ca}$, are present
over a broad range of energies, ranging from X-rays at 67.9\,keV and
78.4\,keV
(from de-excitation of $\AZNucleus{44}{}{Sc}$)
up to $\gamma$-rays at 1157\,keV
(from de-excitation of $\AZNucleus{44}{}{Ca}$).
On average, $\sim 0.93$, $\sim 0.96$, and $\sim 1$ photons are
produced per decay in each of these lines, respectively
\citep{Chen_et_al__2011__NuclearDataSheets__NuclearDataSheetsforA44}.
Lines with energies of 67.9, 78.4, and 1157\,keV for Cas\,A and 67.9
and 78.4 keV for SN\,1987A were observed by many telescopes
\citep{Iyudin_et_al__1994__aap__COMPTELobservationsofTi-44gamma-raylineemissionfromCASA,Iyudin_et_al__1999__AstrophysicalLettersandCommunications__COMPTELAll-SkySurveyin44TILineEmission,Vink_et_al__2001__apjl__Detectionofthe67.9and78.4keVLinesAssociatedwiththeRadioactiveDecayof44TiinCassiopeiaA,Renaud_et_al__2006__apjl__TheSignatureof44TiinCassiopeiaARevealedbyIBISISGRIonINTEGRAL,Grebenev_et_al__2012__nat__Hard-X-rayemissionlinesfromthedecayof44Tiintheremnantofsupernova1987A,Grefenstette_et_al__2014__nat__Asymmetriesincore-collapsesupernovaefrommapsofradioactive44TiinCassiopeiaA,Boggs_et_al__2015__Science__44Tigamma-rayemissionlinesfromSN1987Arevealanasymmetricexplosion,Siegert_et_al__2015__aap__RevisitingINTEGRAL/SPIobservationsof44TifromCassiopeiaA,Tsygankov_et_al__2016__mnras__Galacticsurveyof44TisourceswiththeIBIStelescopeonboardINTEGRAL,Grefenstette_et_al__2017__apj__TheDistributionofRadioactive44TiinCassiopeiaA}.

Historically, the very first detection of \Ti was made by the Compton
scattering telescope COMPTEL on-board the Compton Gamma-Ray
Observatory (CGRO) from the young galactic supernova remnant Cas\,A
\citep{Iyudin_et_al__1994__aap__COMPTELobservationsofTi-44gamma-raylineemissionfromCASA,Iyudin_et_al__1999__AstrophysicalLettersandCommunications__COMPTELAll-SkySurveyin44TILineEmission}
via gamma-ray line emission at 1.157\,MeV.  The first live
confirmation of \Ti (\ie via the observed de-excitation of
$\AZNucleus{44}{}{Sc}$) in the young galactic supernova remnant (SNR)
Cas\,A was obtained
by the low-background spectrometer PDS on-board of Beppo-SAX
\citep{Vink_et_al__2001__apjl__Detectionofthe67.9and78.4keVLinesAssociatedwiththeRadioactiveDecayof44TiinCassiopeiaA}
via the detection of low-energy \Ti decay lines at 68 and 78\,keV.
Later detections of the same low-energy gamma-ray lines at 68 and
78\,keV from \Ti decay were made by ISGRI/IBIS
\citep{Renaud_et_al__2006__apjl__TheSignatureof44TiinCassiopeiaARevealedbyIBISISGRIonINTEGRAL}
on-board of INTEGRAL, and by NuSTAR
\citep{Grefenstette_et_al__2014__nat__Asymmetriesincore-collapsesupernovaefrommapsofradioactive44TiinCassiopeiaA,Grefenstette_et_al__2017__apj__TheDistributionofRadioactive44TiinCassiopeiaA}.
Early all-sky surveys of the \Ti decay line emission were performed by
the COMPTEL Team, first by
\cite{Dupraz_et_al__1997__aap__COMPTELthree-yearsearchforgalacticsourcesof44Tigamma-raylineemissionat1157MeV}
and later by
\cite{Iyudin_et_al__1999__AstrophysicalLettersandCommunications__COMPTELAll-SkySurveyin44TILineEmission}.
Additionally, detection of the 1.15\,MeV gamma-ray line from the
Cas\,A SNR was attempted twice so far by SPI (INTEGRAL), first by
\cite{Martin_et_al__2009__aap__Constraintsonthekinematicsofthe44TiejectaofCassiopeiaAfromINTEGRAL/SPI},
and later by
\cite{Siegert_et_al__2015__aap__RevisitingINTEGRAL/SPIobservationsof44TifromCassiopeiaA}.
The result of an all-sky survey in the \Ti decay lines at 68 and
78\,keV was published recently by
\cite{Tsygankov_et_al__2016__mnras__Galacticsurveyof44TisourceswiththeIBIStelescopeonboardINTEGRAL}
based on ISGRI/IBIS accumulated data.

The conclusion that can be drawn from the measurements reiterated
above is that the \Ti decay line flux detected from the Cas\,A SNR at
1.157\,MeV is larger than that at the 68 and 78\,keV lines, even if
one takes into account the branching ratios of the relevant decays,
and the time of the actual measurements. The reason for this
discrepancy might be related either to the inability to account for
the different detection efficiencies of the telescopes or to the
(astro)-physical properties of the SNR and of its environment
(circumstellar, CSM, and interstellar, ISM, medium).  The dust content
of the CSM and ISM, and the resulting source obscuration could be
especially important in order to better understand why Cas\,A was not
optically observed as a supernova.


It is widely accepted that the large quantities of dust observed in
high-redshift quasars and galaxies
\citep{Priddey_et_al__2008__mnras__Environmentsofz$gt$5quasars:searchingforprotoclustersatsubmillimetrewavelengths,Wang_et_al__2008__apj__ThermalEmissionfromWarmDustintheMostDistantQuasars}
have to be produced rapidly during the galactic evolution.  Massive
stars become significant dust producers in their late evolutionary
stages, reached after a rather brief time on the main sequence,
\citep{Cernuschi_et_al__1967__AnnalesdAstrophysique__Contributiontothetheoryontheformationofcosmicgrains}
and, thus have long been considered the most likely source of the
dust.

Type-II SNe, such as Cas\,A, may also work as dust factories
\citep{Woosley_Weaver__1995__apjs__The_Evolution_and_Explosion_of_Massive_Stars.II.Explosive_Hydrodynamics_and_Nucleosynthesis}.
At present, it is believed that dust is formed in CCSNe in amounts
that are sufficient to explain the dust observed in the distant
universe.  For example,
\cite{McKinnon_et_al__2016__mnras__DustformationinMilkyWay-likegalaxies}
presented a model where roughly two thirds of the dust in
Milky-way-type galaxies is produced by type-II SNe.

This notion, though, until recently was supported only by rather scant
and controversial observations of infrared (IR) satellites like IRAS,
ISO, and Spitzer, which detected mostly a warm dust component of much
smaller mass than predicted by theory. Herschel, operational since
2009
\citep{Pilbratt_et_al__2010__aap__HerschelSpaceObservatory.AnESAfacilityforfar-infraredandsubmillimetreastronomy},
observed SN\,1987A and Cas\,A, and it allowed for a much better
coverage of the far-infrared emission of observed SNRs up to
500\,$\mu\mathrm{m}$. These observations and additional ALMA coverage
of SNRs at 450\,\mum and 870\,\mum
\citep{Zanardo_et_al__2014__apj__SpectralandMorphologicalAnalysisoftheRemnantofSupernova1987AwithALMAandATCA}
strongly support the presence of cold dust in SNRs. Namely, up to
0.8\,$\msol$ of dust is suspected nowadays in SN\,1987A
\citep{Indebetouw_et_al__2014__apjl__DustProductionandParticleAccelerationinSupernova1987ARevealedwithALMA,Matsuura_et_al__2015__apj__AStubbornlyLargeMassofColdDustintheEjectaofSupernova1987A,Wesson_et_al__2015__mnras__ThetimingandlocationofdustformationintheremnantofSN1987A,Dwek_et_al__2015__apj__TheEvolutionofDustMassintheEjectaofSN1987A,Bevan_Barlow__2016__mnras__Modellingsupernovalineprofileasymmetriestodetermineejectadustmasses:SN1987Afromdays714to3604,Matsuura_et_al__2019__mnras}. Similar,
or even larger, amounts of dust are suggested for the Cas\,A SNR by
\cite{Rho_et_al__2008__apj__FreshlyFormedDustintheCassiopeiaASupernovaRemnantasRevealedbytheSpitzerSpaceTelescope,Dunne_et_al__2009__mnras__CassiopeiaA:dustfactoryrevealedviasubmillimetrepolarimetry,Barlow_et_al__2010__aap__AHerschelPACSandSPIREstudyofthedustcontentoftheCassiopeiaAsupernovaremnant,Bevan_et_al__2017__mnras__DustmassesforSN1980KSN1993JandCassiopeiaAfromred-blueemissionlineasymmetries}.

Potential signatures of the formation of dust in the ejecta of CCSNe
are (a) a faster decline of the optical brightness caused by an
increasing extinction due to newly formed dust, (b) an infrared excess
due to the absorption of optical photons and their subsequent
reemission at longer wavelengths, and (c) the absorption of the
red-shifted radiation emitted at the far side of the ejecta by dust
along the line of sight (LoS), thus leading to a relative blue-shift
of the emission lines.

Early definitive detections of dust formation in CCSNe were made for
SN\,1987A as one of the first cases
\citep{Lucy_et_al__1989__IAUColloq.120:StructureandDynamicsoftheInterstellarMedium__DustCondensationintheEjectaofSN1987A,Bouchet_Danziger__1993__aap__InfraredPhotometryandSpectrophotometryofSupernova1987A-PartTwo-1987NOVto1991MARObservations,Meikle_et_al__1993__mnras__Spectroscopyofsupernova1987Aat1-4microns.II-Days377to1114,Wooden_et_al__1993__apjs__AirbornespectrophotometryofSN1987Afrom1.7to12.6microns-Timehistoryofthedustcontinuumandlineemission,Elmhamdi_et_al__2003__mnras__PhotometryandspectroscopyoftheTypeIIPSN1999emfromoutbursttodustformation}.
Assuming a smooth dust distribution, a dust mass of about $10^{-4} \,
\msol$ of dust was found. However in the two well studied cases,
SN\,1987A and SN\,1999em, this assumption was challenged.  A clumpy
rather than homogeneous distribution would correspond to
significantly higher dust masses.

Following early evidence for optically thick clumps by
\cite{Lucy_et_al__1989__IAUColloq.120:StructureandDynamicsoftheInterstellarMedium__DustCondensationintheEjectaofSN1987A,Lucy_et_al__1991__Supernovae__DustCondensationintheEjectaofSupernova1987A-PartTwo},
observations ranging from $\gamma$-rays and X-rays to IR wavelengths by
\cite{Wooden_et_al__1993__apjs__AirbornespectrophotometryofSN1987Afrom1.7to12.6microns-Timehistoryofthedustcontinuumandlineemission}
found significant clumping in the ejecta of SN\,1987A.

More recently,
\cite{Smith_et_al__2008__apj__DustFormationandHeII$lambda$4686EmissionintheDenseShellofthePeculiarTypeIbSupernova2006jc}
found evidence for dust formation 
in the type-Ib SN\,2006jc as early as between 51 and 75 days after the
explosion. They propose that the dust formation occured when the SN
shock wave overtook the dense shell formed by the outburst of the
progenitor just 2yr prior to the SN.
Moreover,
\cite{Gall_et_al__2014__nat__Rapidformationoflargedustgrainsintheluminoussupernova2010jl}
demonstrated the rapid (40 to 240 days after explosion) generation of
dust 
in the dense CSM of the luminous supernova SN\,2010jl.  At later times
(500 to 900 days) their infrared oberservations show that the dust
source is no longer circumstellar matter but SN ejecta,
thus providing a link between the early and late dust mass evolution
in supernovae with dense CSM. The important conclusion drawn from the
dust formation and evolution of SN\,2010jl is that the dust formed
contains quite a high percentage of large ($>1$\,\mum) grains, which
resist destruction during later evolution stages of the SNR.

Studies of grain inclusions found in meteorites delivered an evidence
that at least some SNe produced grains with isotopic distributions
significantly different from those of the Sun and the Earth
\citep[\eg,][]{Clayton_et_al__1997__apss__DustfromSupernovae,Croat_et_al__2003__gca__Structuralchemicalandisotopicmicroanalyticalinvestigationsofgraphitefromsupernovae,Stadermann_et_al__2005__gca__SupernovagraphiteintheNanoSIMS:CarbonoxygenandtitaniumisotopiccompositionsofaspheruleanditsTiCsub-components,Gyngard_et_al__2018__gca__Bonanza:Anextremelylargedustgrainfromasupernova}.

The observations of the early dust formation in SNe quoted above,
together with quite a few new interpretations of the dust content of
the Cas\,A and SN\,1987A SNRs
\citep{DeLooze_et_al__2017__mnras__ThedustmassinCassiopeiaAfromaspatiallyresolvedHerschelanalysis,Bevan_et_al__2017__mnras__DustmassesforSN1980KSN1993JandCassiopeiaAfromred-blueemissionlineasymmetries}
inspired us to propose a new interpretation of the \Ti fluxes measured
for Cas\,A at different energies and by different instruments.  Below
we consider consequences of dust presence in the CSM of SNR, and in
the ISM along the LoS to SNR. The Cas\,A SNRis our primary object of
interest, because it is well studied with respect to dust presence as
well as to \Ti lines measurements.

\section{Cas A dust and \Ti content}
\label{Sek:CasAdustTi}

\begin{figure}
  \centering
  \includegraphics[width=\linewidth]{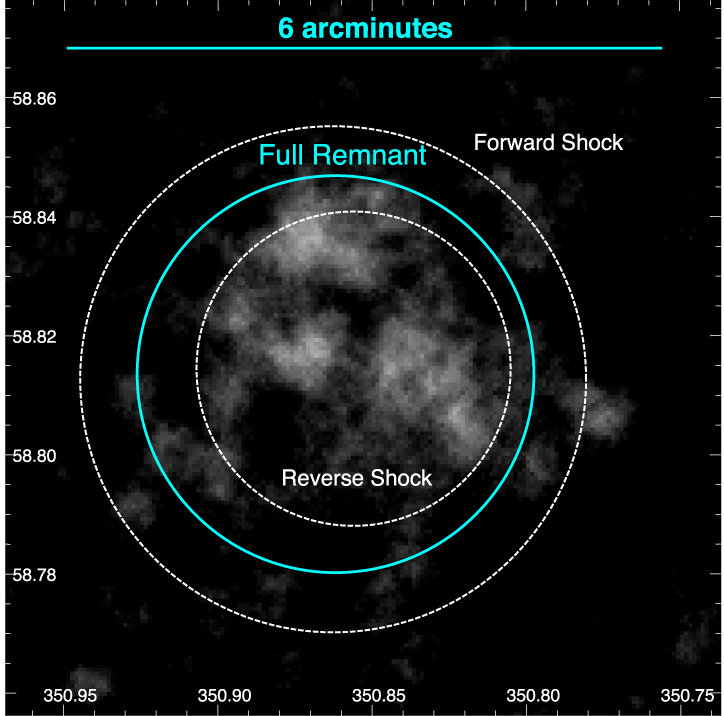}
  \caption{
    NuSTAR measurements of the distribution of \Ti clumps in Cas\,A. 
    Circles are selfexplanatory. Adapted from
    \protect\cite{Grefenstette_et_al__2017__apj__TheDistributionofRadioactive44TiinCassiopeiaA}.
  }
  \label{Fig:CasA-1}
\end{figure}

\begin{figure}
  \centering
  \includegraphics[width=\linewidth]{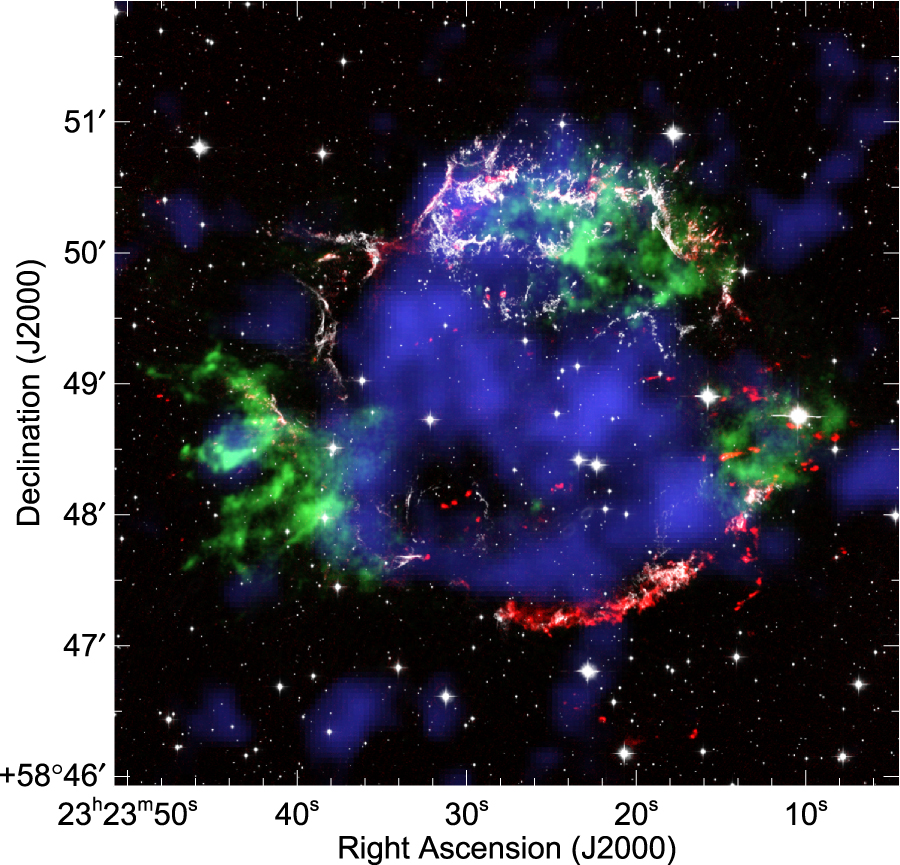}
  \caption{
    Four-colour composite image of Cas\,A \protect\citep[Fig.~12
    of][]{Lee_et_al__2017__apj__Near-InfraredKnotsandDenseFeEjectaintheCassiopeiaASupernovaRemnant}:
    in red the [Fe II] 1.644\,\mum narrow-band image (their Fig.~1),
    in green the Chandra Fe K-shell (6.52 - 6.94\,keV) image of
    \protect\cite{Hwang_et_al__2004__apjl__AMillionSecondChandraViewofCassiopeiaA},
    in blue the NuSTAR hard X-ray \Ti (67.9 and 78.4\,keV) image of
    \protect\cite{Grefenstette_et_al__2014__nat__Asymmetriesincore-collapsesupernovaefrommapsofradioactive44TiinCassiopeiaA},
    and in white the HST ACS/WFC F850LP image of
    \protect\cite{Fesen_et_al__2006__apj__DiscoveryofOutlyingHigh-VelocityOxygen-RichEjectainCassiopeiaA}.
  }
  \label{Fig:CasA-2}
\end{figure}

\begin{figure}
  \centering
  \includegraphics[width=\linewidth]{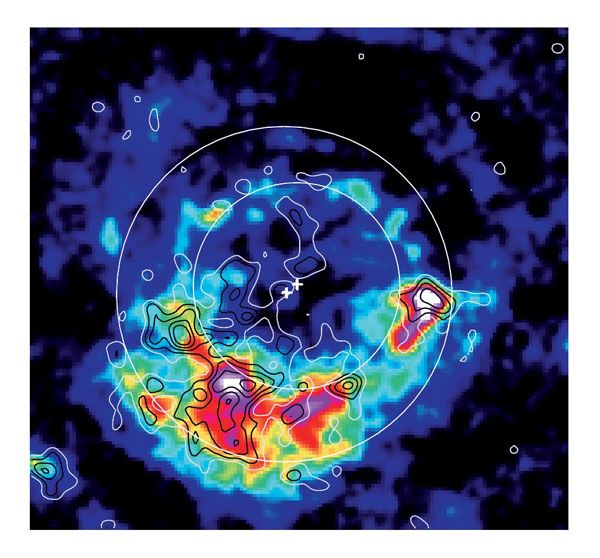}
  \caption{
    Dust distribution in the Cas\,A SNR based on 
    $850 \, \mu \mathrm{m}$ IR measurements \protect\citep[adapted
    from][]{Dunne_et_al__2003__nat__TypeIIsupernovaeasasignificantsourceofinterstellardust}. North
    is up, and east is to the left.  Circles correspond to the reverse
    ($95 \pm 10$ arcsec) and forward ($153 \pm 12$ arcsec) shocks. The
    image size is 8.4' by 7.8'. The synchrotron emission was
    subtracted using the 83\,GHz image by
    \protect\cite{Wright_et_al__1999__apj__TheSupernovaRemnantCassiopeiaAatMillimeterWavelengths}.
  }
  \label{Fig:CasA-1b}
\end{figure}

\begin{figure}
  \centering
  \includegraphics[width=\linewidth]{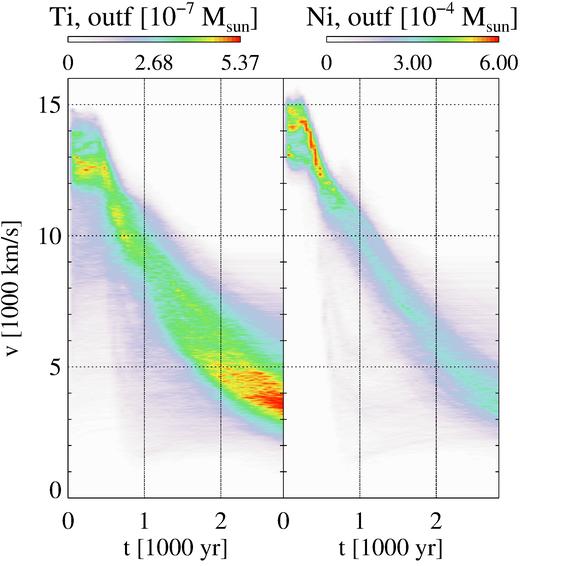}
  \caption{
    Velocity spectra of \Ti and $\AZNucleus{56}{}{Fe}$ in the outflows
    of the asymmetric explosion model 25A from
    \protect\cite{Obergaulinger_et_al__2014__mnras__HydrodynamicsimulationsoftheinteractionofsupernovashockwaveswithaclumpyenvironmentthecaseoftheRXJ0852.0-4622(VelaJr)supernovaremnant}
    at different times after the explosion.
  }
  \label{Fig:Simulation}
\end{figure}

\begin{table*}
  \centering
  \begin{tabular}{|l|lll|lll|}
    \hline
    No & $ f (67.86 \, \mathrm{keV}) $ &  $ f (78.4 \, \mathrm{keV}) $
    &  $ f (1157 \, \mathrm{keV}) $ & date(s) & Intrument & comments
    \\
    \hline
    1 & -- & -- & $3.3 \pm 0.6$ & 1991--1997 & COMPTEL & VP0.1-617.1
    \\
    2 & $2.1 \pm 0.7$ & $2.1 \pm 0.7$ & -- & May, June 2001 & BeppoSAX
    & 12-300 keV pl continuum
    \\
    3 & $1.9 \pm 0.9 $ & $1.9 \pm 0.9$ & -- & May, June 2001 &
    BeppoSAX & 30-100 keV pl continuum
    \\
    4 & $2.3 \pm 0.5$ & $2.3 \pm 0.5$ & -- & 2003-2005 & IBIS (INT) & 
    \\
    5 & $1.51 \pm 0.31$ & -- & -- & Aug 2012, Jun 2013 & NuSTAR 1.2 Ms
    & 67.8 line redshifted
    \\
    6 & -- & $2.1 \pm 0.4$ & $3.1 \pm 1.2$ & 2003-2014 & SPI (INT) &
    \\
    7 & $1.84 \pm 0.25$ & -- & -- & Aug 2012--Dec 2013 & NuSTAR & 2.4
    Ms exposure
    \\ \hline
  \end{tabular}
  \caption{
    \Ti lines flux measurements for the Cas\,A remnant.  The fluxes (second
    to fourth columns) are given in units of $10^{-5}\, \cm^{-2} \mathrm{s}^{-1}$
  }
  \label{Tab:Tifluxes}
\end{table*}

\begin{table*}
  \centering
  \begin{tabular}{|l|lllll|}
    \hline
    Dust & Density 
    &   Attenuation 
    &   Attenuation 
    &   Attenuation 
    &   Attenuation 
    \\ 
    composition & $[\mathrm{g} \mathrm{cm}^{-2}]$ 
    &   at 4.5 keV
    &   at 67.78 keV
    &   at 78.36 keV
    &   at 1157 keV
    \\ \hline
    SiC & & 236.8 & 0.2391& 0.2078 & 0.05917
    \\ 
    $\mathrm{Mg} \mathrm{Al}_2 \mathrm{O}_4$
    & & 163.6 & 0.2095 & 0.1879 & 0.05833
    \\ 
    $\mathrm{Mg}_2 \mathrm{Si} \mathrm{O}_4$
    & 3.32 & 168.1 & 0.2131 & 0.1908 & 0.05899
    \\ 
    Fe
    & 7.95 & 186.2 & 0.08800 & 0.6234 & 0.05565
    \\ 
    $\mathrm{Fe}_3 \mathrm{O}_4$
    & 5.25 & 150.8 & 0.6864 & 0.4979 & 0.05664
    \\ 
    FeS
    & 4.87 & 284.4 & 0.6785 & 0.4929 & 0.05696
    \\ \hline
  \end{tabular}
  \caption{
    Attenuation coefficients for different dust compositions at
    photon energies of 4.5\,keV, 68\,keV, 78\,keV, and 1.157\,MeV in units
    of $\cm^{2} / \mathrm{g}$.
  }
  \label{Tab:AttCoeff}
\end{table*}

It is well known that the presence of dust grains around X-ray sources
produces X-ray halos around bright sources due to small-angle X-ray
scattering
\citep{Mauche_Gorenstein__1986__apj__MeasurementsofX-rayscatteringfrominterstellargrains,Predehl_Schmitt__1995__aap__X-rayingtheinterstellarmedium:ROSATobservationsofdustscatteringhalos,Valencic_Smith__2015__apj__InterstellarDustPropertiesfromaSurveyofX-RayHalos}.
Below we will demonstrate that the presence of dust in SNRs might
affect the \Ti line measurements at hard X-ray and $\gamma$-rays using
the Cas\,A SNR as an example.

The dust mass in the Cas\,A SNR was recently re-derived based on a
spatially resolved Herschel analysis
\citep{DeLooze_et_al__2017__mnras__ThedustmassinCassiopeiaAfromaspatiallyresolvedHerschelanalysis}
by far-infrared (Herschel) and submillimeter (ALMA) measurements of
the emission spectra of the cold dust in this young galactic SNR. The
new dust mass estimate based on the far-infrared and submillimeter SED
of this SNR, together with the dust mass estimate based on the
analysis of the shape of oxygen lines by
\cite{Bevan_et_al__2017__mnras__DustmassesforSN1980KSN1993JandCassiopeiaAfromred-blueemissionlineasymmetries},
leads to a total dust mass estimate of $\sim 1\, \msol$.  This dust
mass is much larger than what was evaluated previously on the basis of
measurements at shorter wavelengths
\citep{Rho_et_al__2008__apj__FreshlyFormedDustintheCassiopeiaASupernovaRemnantasRevealedbytheSpitzerSpaceTelescope,Barlow_et_al__2010__aap__AHerschelPACSandSPIREstudyofthedustcontentoftheCassiopeiaAsupernovaremnant,Arendt_et_al__2014__apj__InterstellarandEjectaDustintheCasASupernovaRemnant}.

Apart from the total dust mass estimate, the spatial distribution of
the dust is important to understand, too.  We note that warm dust was
detected mainly inside of the SNR reverse shock (RS).  But it is quite
plausible to expect the presence of a much higher mass of cold dust
inside, or near, the 
cold dense shell (CDS)
between the RS and the forward shock (FS), which is likely also very
clumpy. This conclusion follows from observations at
850\,$\mu \mathrm{m}$ by
\cite{Dunne_et_al__2009__mnras__CassiopeiaA:dustfactoryrevealedviasubmillimetrepolarimetry},
(see also \figref{Fig:CasA-1b}, and the discussion in
\cite{Dunne_et_al__2009__mnras__CassiopeiaA:dustfactoryrevealedviasubmillimetrepolarimetry}
on polarised dust measured in Cas\,A. It is also supported by the
results of
\cite{Bevan_et_al__2017__mnras__DustmassesforSN1980KSN1993JandCassiopeiaAfromred-blueemissionlineasymmetries}
on the asymmetry of the [OIII] line observed from the region between
RS and CDS in the Cas\,A SNR.

The observations by
\cite{Lee_et_al__2015__apj__Near-infraredExtinctionduetoCoolSupernovaDustinCassiopeiaA}
of excessive and varying extinction along different LoS to the Cas\,A
SNR, measured by using [FeII] line emission at 1.26\,\mum and
1.64\,\mum, and their lines flux ratio, may give an indication of the
opacity of this region as well. Using these results, as well as
additional spectroscopic results for 63 [FeII] line-emitting knots,
\cite{Lee_et_al__2017__apj__Near-InfraredKnotsandDenseFeEjectaintheCassiopeiaASupernovaRemnant}
concluded that some LoS crossing the shell that is bright in [FeII]
(\figref{Fig:CasA-2} and Fig. 10 (left) of
\cite{Lee_et_al__2017__apj__Near-InfraredKnotsandDenseFeEjectaintheCassiopeiaASupernovaRemnant})
might correspond to column densities of Fe ranging from
$N_{Fe} \sim (2-5) \times 10^{18}\, \cm^{-2}$ in the gas phase up to
$N_{Fe} \sim 5 \times 10^{21}\, \cm^{-2}$ in the dusty clumps that are
filling the shell beyond the RS with reasonably small filling factors.

These results of the dust distribution and the related extinction
measured by using [FeII] line emission
\cite{Lee_et_al__2015__apj__Near-infraredExtinctionduetoCoolSupernovaDustinCassiopeiaA},
together with extinction maps for the field in and around Cas\,A gives
average values of $A_{\mathrm{V}} = 6- 8$\,mag and up to
$A_{\mathrm{V}} = 15$\,mag towards Cas\,A
\citep{Eriksen_et_al__2009__apj__TheReddeningTowardCassiopeiaAsSupernova:Constrainingthe56NiYield,Lee_et_al__2015__apj__Near-infraredExtinctionduetoCoolSupernovaDustinCassiopeiaA,DeLooze_et_al__2017__mnras__ThedustmassinCassiopeiaAfromaspatiallyresolvedHerschelanalysis}.
In a similar fashion, we consider the ratios of the measured \Ti flux
from Cas\,A at different energies as a consequence of the presence of
dust in Cas\,A.

\tabref{Tab:Tifluxes} provides an overview of the published
measurements of \Ti line emission from Cas\,A obtained by different
instruments.

In addition to the measured fluxes, it is important to consider also
the velocity distribution of Ti in Cas\,A.  IBIS/ISGRI measurements
resulted in an upper limit for the \Ti velocity $< 14 000 \, \kms$
\citep{Renaud_et_al__2006__apjl__TheSignatureof44TiinCassiopeiaARevealedbyIBISISGRIonINTEGRAL}.

COMPTEL was an imaging instrument, but its angular resolution at
1.157\,MeV was $\sim 2$ degrees only, i.e., much larger than the
angular size of Cas\,A. Therefore, it was only possible to estimate
the mean velocity of the \Ti bulk motion in the SNR as
$v(^{44}\mathrm{Ti}) = (7200 \pm 2900) \, \kms$ from the COMPTEL
measurements of the 1.157 MeV line energy distribution
\citep{Iyudin_et_al__1997__TheTransparentUniverse__CasAintheLightofthe44Ti1.15MeVGamma-RayLineEmission}.
Recent NuSTAR results are of utmost importance for the intended
comparison because NuSTAR is currently the only imaging experiment
that is capable to derive the 3D spatial distribution of \Ti in the
Cas\,A SNR
\citep{Grefenstette_et_al__2014__nat__Asymmetriesincore-collapsesupernovaefrommapsofradioactive44TiinCassiopeiaA,Grefenstette_et_al__2017__apj__TheDistributionofRadioactive44TiinCassiopeiaA}.

The main conclusions from NuSTAR measurements are that radioactive \Ti
in Cas\,A is moving with a mean velocity of $\sim 5000 \, \kms$, and
its spatial distribution inside the SNR is not consistent with that of
\Ni
\citep{Grefenstette_et_al__2014__nat__Asymmetriesincore-collapsesupernovaefrommapsofradioactive44TiinCassiopeiaA,Grefenstette_et_al__2017__apj__TheDistributionofRadioactive44TiinCassiopeiaA},
even though there is some correspondence of \Ti and Fe K-$\alpha$
emission at the Northwest and Southwest parts of the SNR
\citep{Grefenstette_et_al__2017__apj__TheDistributionofRadioactive44TiinCassiopeiaA}.

Eight out of 11 significantly detected clumps of \Ti measured by
NuSTAR are projected inside the reverse shock
\citep{Grefenstette_et_al__2017__apj__TheDistributionofRadioactive44TiinCassiopeiaA}
The exceptions are three clumps, one of which moves with a LoS
velocity of $\sim 8000 \, \kms$ towards the observer.
At the same time, in agreement with the velocity of \Ti measured by
COMPTEL, \Ti clumps must be spread out up to radii of $\sim 150''$,
which correspond to the position of the Cas\,A forward shock.
Considering these differencies of the measured velocities, we
speculate that NuSTAR has not detected any fast component of \Ti that
moves close to the plane of the sky beyond the reverse shock for
unknown reasons, and, therefore, it has measured only an, albeit
large, part of the total flux of decaying \Ti that is present today in
Cas\,A.

We also note that simulation results of an asymmetric explosion of a
$25 \, \msol$ star point towards much higher velocities of \Ti at the
age of the Cas\,A SNR, see Fig.~5 of
\cite{Obergaulinger_et_al__2014__mnras__HydrodynamicsimulationsoftheinteractionofsupernovashockwaveswithaclumpyenvironmentthecaseoftheRXJ0852.0-4622(VelaJr)supernovaremnant},
and our \figref{Fig:Simulation}. Looking at the velocity distributions
of \Ti and Fe (Ni) at $\sim 320$\,yrs after the SN explosion we note a
substantial overlap of the distributions for both elements, with the
amount of Fe (Ni) exceeding $\sim 10^3$ times that of \Ti at the same
velocities.

Coming back to the flux values given in \tabref{Tab:Tifluxes}, it is
tempting to explain the deficiency of the measured fluxes at energies
$\sim 68$\,keV and $\sim 78$\,keV compared to the fluxes measured at
1.157\,MeV as a consequence of an additional component of the flux at
1.157\,MeV, which is generated by the interaction of low-energy cosmic
rays (LECR) with abundant circumstellar species like Fe, Mn, and Cr in
the environment of Cas\,A.
\cite{Siegert_et_al__2015__aap__RevisitingINTEGRAL/SPIobservationsof44TifromCassiopeiaA}
attempted to interpret these discrepant line fluxes of \Ti at
different energies exactly in this way. However, considering that
little or nothing is known about the LECR fluxes in the Cas\,A SNR as
well as uncertainties in the generally small cross sections for the
relevant reactions to produce either \Ti or excited
$\AZNucleus{44}{}{Ca}^{*}$
\cite{Silberberg_et_al__1998__apj__UpdatedPartialCrossSectionsofProton-NucleusReactions,Tsao_et_al__1998__apj__PartialCrossSectionsofNucleus-NucleusReactions},
the validity of such an interpretation of the discrepant measurements
of \Ti fluxes at 68 and 78\,keV and at 1.157\,MeV by different
instruments as well as by the same instrument (SPI) may be
doubted. Additionally, if one assumes that the production of the
1.157\,MeV line emission is supported in Cas\,A by the enhanced flux
of LECRs, then immediately a problem arises related to the absence of
other lines emission from much more abundant elements like oxygen,
carbon and nitrogen with very large, well measured cross-sections for
excitation. Indeed, simulations of the excitation emission lines
produced by LECRs in Cas\,A have shown that lines at 4.43, 6.13, 6.9,
and 7.1\,MeV are expected to be strongest (Summa et al. 2011) for the
expected spectrum of LECR (Berezhko et al. 2003) that are interacting
with the CSM near the Cas\,A SNR. However, because no gamma-ray line
emission at 4.43, 6.13, 6.9, and 7.1\,MeV is observed from the Cas\,A
SNR, we consider this explanation of the excitation origin of the high
1.157\,MeV \Ti line flux compared to that at 68 and 78\,keV as very
problematic.

In our opinion, a more appealing explanation for the observed
differences of the measured line fluxes in the emission of \Ti is the
dependence of the line flux measurements at different energies on the
SNR environment and on the gas-to-dust ratio in the ejecta.  Coming
back to the recent measurements of the dust content of Cas\,A and of
the ISM along LoS towards Cas\,A discussed above, it is fair to assume
that a part of the emitted photons at energies of 68 and 78\,keV are
absorbed or scattered, likely by the dust particles, out of the sight
lines. Indeed, looking at the values of the attenuation coefficients
for photon interaction at energies of 68-78\,keV and 1.157\,MeV for a
dust composition taken as \MgSiO (\tabref{Tab:AttCoeff}), one
immediately apprehends that the attenuation of photons with energies
of 68-78\,keV is notably larger than that of photons with an energy of
1.157\,MeV. An attenuation coefficient is generally derived from the
depletion of a photon beam while passing through the ISM or a cloud
with known abundances. For photons with energies of $\sim 68$ and
$\sim 78$\,keV, photoeffect and Compton scattering are the most
important processes removing photons from the beam, while for
1.157\,MeV energies only Compton scattering has to be considered.

Let us have a more quantitative look at the problem. For this we will
use \tabref{Tab:AttCoeff} which contains attenuation coefficients for
different dust compositions at photon energies of $\sim 68$\,keV,
$\sim 78$\,keV, and 1.157\,MeV, and the line flux measurements of
\tabref{Tab:Tifluxes}.  To compare the measured fluxes of \Ti at
different line energies we re-calculate the measured fluxes to the
chosen time period after the SN explosion. Because the latest
measurements of the \Ti line flux at 67.9\,keV by NuSTAR
\citep{Grefenstette_et_al__2017__apj__TheDistributionofRadioactive44TiinCassiopeiaA}
were performed from August 2012 to December 2013 or at 2013.3, we take
the period of 2013.3 as the mean time of actual measurement by NuSTAR
and will re-calculate all other measured fluxes to this date taking
the \Ti decay into account.
Using this conversion of the measured fluxes we come to weighted means
of the \Ti measured fluxes at 1.157\,MeV and 67.9\,keV as follows:
\begin{eqnarray}
  \label{Eq:TiFlux1157}
  \langle F_{1.157 \, \MeV}  \rangle
  & = & 
  \left(
    2.70 \pm 0.43
  \right)
  \times 10^{-5}
  \, \frac{ \mathrm{photons} }{ \sek \, \cm^2},
  \\
  \label{Eq:TiFlux00679}
  \langle F_{67.9 \, \keV}  \rangle
  & = & 
  \left(
    1.96 \pm 0.22
  \right)
  \times 10^{-5}
  \, \frac{ \mathrm{photons} }{ \sek \, \cm^2},
\end{eqnarray}

These are significantly different values for the \Ti line fluxes,
though with large error bars. If one takes into account branching of
lines, the derived final ratio of line fluxes at 67.9\,keV and
1.157\,MeV is equal to:
\begin{equation}
  \label{Gl:fluxes3}
  \frac{ \langle F_{67.9 \, \mathrm{keV}} \rangle} 
  {\langle F_{1.157 \, \mathrm{MeV}} \rangle} = 0.78^{+ 0.26}_{ - 0.18} .
\end{equation}

Such a ratio can be achieved if the line at the smaller energy
(68\,keV) was attenuated by a dust clump composed of \MgSiO of
$1.18 \, \grm \cm^{-2}$ thickness along the line of sight, placed
somewhere between 2.5\,pc and $\sim 4$\,pc radial distance from the
Cas\,A explosion centre, \eg at larger radii than the reverse
shock. For a dust clump composed of \FeO, the same absorption effect
can be achieved with an attenuation length of about
$0.36 \, \grm \cm^{-2}$ along the LoS towards the \Ti clumps. We note
that quite some amount of Fe is present between the RS and the FS,
according to measurements of [FeII] line emission by
\cite{Smith_et_al__2009__apj__SpitzerSpectralMappingofSupernovaRemnantCassiopeiaa},
as shown in their Fig.~1, or by
\cite{Lee_et_al__2017__apj__Near-InfraredKnotsandDenseFeEjectaintheCassiopeiaASupernovaRemnant}.

Interestingly, about the same quantity of absorbing matter containing
Fe along the LoS to Cas\,A was derived by
\cite{Lee_et_al__2015__apj__Near-infraredExtinctionduetoCoolSupernovaDustinCassiopeiaA}
from their analysis of the IR absorption along the LoS towards Cas\,A.
From the relative positions of \Ti clumps detected by NuSTAR
\citep{Grefenstette_et_al__2014__nat__Asymmetriesincore-collapsesupernovaefrommapsofradioactive44TiinCassiopeiaA,Grefenstette_et_al__2017__apj__TheDistributionofRadioactive44TiinCassiopeiaA},
and the dust distribution and the polarization that was detected in
Cas\,A SNR by
\cite{Dunne_et_al__2009__mnras__CassiopeiaA:dustfactoryrevealedviasubmillimetrepolarimetry},
see Fig.~1 and 2, the interpretation of some part of the \Ti being
placed somewhere at distances between the RS and the FS is quite in
order. If this scenario holds, then it is reasonable to assume that
some fraction of the 68 and 78\,keV line emission of \Ti was absorbed
or scattered off the LoS by matter located along the LoS towards the
\Ti clumps ejected with velocities higher than $\sim 5000$\,km/s
during the SN explosion.

Note that absorption might happen for the \Ti clumps observed by
NuSTAR as well. However, this absorption could be smaller due to the
dust filling factor of the absorbing shell, or due to the smaller dust
density or/and favourable dust composition along appropriate lines of
sight.  Actually, we think that 
apart from clump 20
\cite{Grefenstette_et_al__2017__apj__TheDistributionofRadioactive44TiinCassiopeiaA}
\Ti could be detected in Cas\,A by NuSTAR beyond the RS, in the sky
projection. However, this
will demand a much longer exposure
because the signal to noise (S/N) ratio of the 68 and 78\,keV lines
fluxes will likely be much smaller outside the RS
due to the dilution of a smaller amount of high-velocity \Ti in the
dense shell between the RS and the FS.

\begin{figure}
  \centering
  \includegraphics[width=\linewidth]{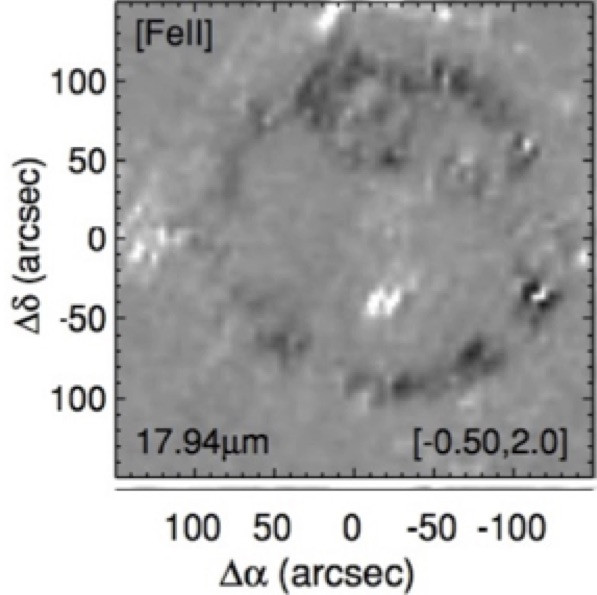}
  \caption{
    Surface brightness map in the [Fe II] line at 17.94\,\mum. The image
    size is $300'' \times 300''$ with offset position as in Fig.~1 of
    \protect\cite{Smith_et_al__2009__apj__SpitzerSpectralMappingofSupernovaRemnantCassiopeiaa}. The
    integrated, continuum-subtracted line surface brightness is square
    root scaled between the low and high thresholds specified in brackets
    (in units of
    $10^{-7} \, \mathrm{W} \mathrm{m}^{-2} \mathrm{sr}^{-1}$).
  }
  \label{Fig:FeII}
\end{figure}

\section{Discussion}
\label{Sek:Discussion}

The real amount of dust in the Cas\,A SNR, its spatial distribution,
and the reddening towards Cas\,A remain unsettled problems at present,
see
\cite{Dunne_et_al__2003__nat__TypeIIsupernovaeasasignificantsourceofinterstellardust,Krause_et_al__2004__nat__NocolddustwithinthesupernovaremnantCassiopeiaA,Dwek__2004__apj__TheDetectionofColdDustinCassiopeiaA:EvidencefortheFormationofMetallicNeedlesintheEjecta,Dunne_et_al__2009__mnras__CassiopeiaA:dustfactoryrevealedviasubmillimetrepolarimetry,Eriksen_et_al__2009__apj__TheReddeningTowardCassiopeiaAsSupernova:Constrainingthe56NiYield,Sibthorpe_et_al__2010__apj__AKARIandBLASTObservationsoftheCassiopeiaASupernovaRemnantandSurroundingInterstellarMedium,Lee_et_al__2015__apj__Near-infraredExtinctionduetoCoolSupernovaDustinCassiopeiaA,Bevan_et_al__2017__mnras__DustmassesforSN1980KSN1993JandCassiopeiaAfromred-blueemissionlineasymmetries,DeLooze_et_al__2017__mnras__ThedustmassinCassiopeiaAfromaspatiallyresolvedHerschelanalysis}.
The amount of dust inside and towards the Cas\,A SNR is debatable, but
the usual argument that the amount of dust produced after the
explosion cannot exceed the total mass of the SN ejecta can be false,
because it ignores the possible contribution from the condensible
material in the pre-supernova wind and the possible flaring history of
the progenitor. This contribution may include the mass ejected by the
supernova progenitor in the form of stellar wind and/or of the ejected
matter by the star during the so-called supernova ``impostor''
phase. In the case of the Cas\,A SNR one has to take seriously the
observation of the so-called ``Guest-Star'' near the position of
Cas\,A
\cite{Park_et_al__2016__JournalofKoreanAstronomicalSociety__TheKorean1592-1593RecordofaGuestStarAnImpostoroftheCassiopeiaASupernova}
made by Korean astronomers in 1592 - 1593\,AD.  This ``Guest-Star''
may have been an LBV-type outburst of the Cas\,A progenitor, similar
to that $\eta$\,Car in 1843
\cite{Smith_Owocki__2006__apjl__OntheRoleofContinuum-drivenEruptionsintheEvolutionofVeryMassiveStarsandPopulationIIIStars,Smith__2008__nat__Ablastwavefromthe1843eruptionofetaCarinae},
but likely on a smaller scale.  In that case, the Cas\,A
pre-supernova would have ejected considerable mass and over the
following $\sim 80$ years produced notable amounts of dust until the
final explosion around 1671\,AD
\cite{Thorstensen_et_al__2001__aj__TheExpansionCenterandDynamicalAgeoftheGalacticSupernovaRemnantCassiopeiaA,Soria__2013__pasa__OnCasACassiniCometsandKingCharles}.
The dust produced during this ``impostor'' phase would then act as a
``veil'' for the final explosion.

It is well observed that some of the superluminous supernovae are
bright because their ejecta are interacting with the dense CSM
\citep{Chugai_Danziger__1994__mnras__Supernova1988Z-Low-MassEjectaCollidingwiththeClumpyWind,Smith_et_al__2010__apj__SpectralEvolutionoftheExtraordinaryTypeIInSupernova2006gy,Smith__2017__PTRSA__InteractingSupernovaeTypesIInandIbn}.
Moreover, some parts of the ejected nucleosynthesis products condense
into dust grains quite early after the explosion (on time scales of
$\sim$ 40 - 240\,days). Another SNR, produced by SN\,1987A, for which
an emission of \Ti at 68 and 78\,keV was detected
\cite{Boggs_et_al__2015__Science__44Tigamma-rayemissionlinesfromSN1987Arevealanasymmetricexplosion},
is suspected to have at $\sim 9500$\,days after the explosion
$\sim 1 \msol$ of dust
\citep{Wesson_et_al__2015__mnras__ThetimingandlocationofdustformationintheremnantofSN1987A,Bevan_et_al__2017__mnras__DustmassesforSN1980KSN1993JandCassiopeiaAfromred-blueemissionlineasymmetries}.
For some events, like SN\,2010jl, dust grains produced at these early
stages of the ejecta expansion are quite large, up to
4.2\,$\mu \mathrm{m}$
\citep{Gall_et_al__2014__nat__Rapidformationoflargedustgrainsintheluminoussupernova2010jl}.
The reason for such an early growth of dust in the ejecta may be
related to the presence of a Cold Dense Shell (CDS), formed before the
SN explosion, like in the case of SN\,2010jl
\cite{Jencson_et_al__2016__mnras__OpticalobservationsoftheluminousTypeIInSupernova2010jlforover900d}.
Interestingly, [FeII] and a large part of the X-ray Fe emission in
Cas\,A SNR are found between RS and FS (see \figref{Fig:CasA-2} and
\figref{Fig:FeII}), where a CDS might exist.

Noteworthy, Ti has the highest depletion probability among other
elements in the Milky Way
\citep{Jenkins__2009__apj__AUnifiedRepresentationofGas-PhaseElementDepletionsintheInterstellarMedium},
and Fe is second to Ti in this respect. Hence, we speculate that some
part of the \Ti is incorporated into dust grains which might also
contain some Fe. However, such grains will consist not of pure Fe or
Ti agglomerates, but will contain much larger masses of graphite, or
of SiC; see, for example, Figure 9 in the paper by
\cite{Stadermann_et_al__2005__gca__SupernovagraphiteintheNanoSIMS:CarbonoxygenandtitaniumisotopiccompositionsofaspheruleanditsTiCsub-components}
showing an image of a graphite slice, or Figure 2 and the discussion
section in the paper by
\cite{Croat_et_al__2003__gca__Structuralchemicalandisotopicmicroanalyticalinvestigationsofgraphitefromsupernovae}.
The phase condensation of first TiC, with later co-added Fe, and
finally covered by graphite, inferred by
\cite{Croat_et_al__2003__gca__Structuralchemicalandisotopicmicroanalyticalinvestigationsofgraphitefromsupernovae},
implies that \Ti is located in the central bulk region of dust
grains. This conclusion is supported by the radial number distribution
of TiC that is decreasing with increasing distance from the graphite
grain centre
\cite{Croat_et_al__2003__gca__Structuralchemicalandisotopicmicroanalyticalinvestigationsofgraphitefromsupernovae}.
The addition of Fe on top of a pebble of TiC implies that the decay
emission of \Ti, if it is present in this particular dust grain, will
have to penetrate the absorbing matter of the dust grain, including
apart from graphite, or SiC, also some amount of Fe, which is much
more effective absorber than lighter elements like C and Si (see
\tabref{Tab:AttCoeff}).

Considering Ti depletion into the dust fraction together with steadily
updated values of the dust mass measured in the Cas\,A SNR, one has to
pose a question: what influence might the CSM and ISM dust have for
the measurements of \Ti line fluxes from Cas\,A at 68-78\,keV and
1.157\,MeV energies and on the observability of the Cas\,A SN?

Let us reiterate here that Cas\,A has a known problem with not being
observed as a SN (despite the claim of being seen by the Astronomer
Royal Sir John Flamsteed
\citep{Ashworth__1980__JournalfortheHistoryofAstronomy__AProbableFlamsteedObservationoftheCassiopeiaASupernova}).
This problem can be solved or, at least, eased a lot, by the
assumption of the progenitor being shadowed by the dust veil produced
by the SNe impostor of 1592-1593, in agreement with the Korean record
of a Guest Star
\citep{Park_et_al__2016__JournalofKoreanAstronomicalSociety__TheKorean1592-1593RecordofaGuestStarAnImpostoroftheCassiopeiaASupernova},
and with the hypothesis put forward by
\cite{Predehl_Schmitt__1995__aap__X-rayingtheinterstellarmedium:ROSATobservationsofdustscatteringhalos}
and
\cite{Hartmann_et_al__1997__NuclearPhysicsA__OnFlamsteedssupernovaCasA}.

Accepting the role of the ``impostor'' phase in Cas\,A for the
non-detection of the SN as well as the physical mechanism that leads
to an ``impostor'' phase during He-C and Ne shell burning, then it is
possible to accept also that dust grains produced after the final
explosion of the Cas\,A progenitor will show signatures of unusually
heavy SiC with $\AZNucleus{29,30}{}{Si}$ enrichment inside of
graphites, i.e., similar to dust grains found inside of the Murchison
meteorite, with accompanying $\AZNucleus{12}{}{C}$ enrichment, like in
the case discussed by
\cite{Croat_et_al__2010__aj__Unusual2930Si-richSiCsofMassiveStarOriginFoundWithinGraphitesfromtheMurchisonMeteorite}.

According to
\cite{Dwek_et_al__2015__apj__TheEvolutionofDustMassintheEjectaofSN1987A,Wesson_et_al__2015__mnras__ThetimingandlocationofdustformationintheremnantofSN1987A,Sluder_et_al__2018__MNRAS__Molecularnucleationtheoryofdustformationincore-collapsesupernovaeappliedtoSN1987A},
after 104 days of SNR evolution the dust might cool down below
$30\, \mathrm{K}$. Such low temperatures require measurements at long
wavelengths (far-infrared, mm) to study the dust distribution.  It is
also noteworthy that dust grains tend to have quite large sizes at
this stage of SNR evolution
\citep{Dwek_et_al__2015__apj__TheEvolutionofDustMassintheEjectaofSN1987A,Wesson_et_al__2015__mnras__ThetimingandlocationofdustformationintheremnantofSN1987A,Sluder_et_al__2018__MNRAS__Molecularnucleationtheoryofdustformationincore-collapsesupernovaeappliedtoSN1987A},
sometimes up to $\sim 25$\,\mum, like that for the grain termed
Bonanza
\citep{Gyngard_et_al__2018__gca__Bonanza:Anextremelylargedustgrainfromasupernova}.

Considering the broadly accepted extinction towards Cas\,A of
$A_{\mathrm{V}} \sim 6$ to 8 magnitudes based on measurements and
analyses by
\cite{Eriksen_et_al__2009__apj__TheReddeningTowardCassiopeiaAsSupernova:Constrainingthe56NiYield},
together with an additional visual extinction of $\sim 2.8$ magnitudes
produced by the Cas\,A ``impostor'' ejecta in the years 1592-1593, and
by taking the subsequent dilution of this impostor ejecta until the
NuStar observation into account
might solve the puzzle of an otherwise normal SN\,IIb type event not
being contemporaneously observed by astronomers in Europe.

If a notable amount of radioactive \Ti is hidden in dust grains as we
suggested above, then it is difficult to expect, at least for that
part of a SNR where the dust is abundant, that the ionisation state of
\Ti is similar to that of Fe as was proposed by
\cite{Laming_Hwang__2007__RevistaMexicanadeAstronomiayAstrofisicaConferenceSeries__TheCassiopeiaASupernovaRemnantinX-Rays}.
It would be similarly difficult to expect the stripped Ti to quench
the decay of \Ti as was suggested by
\cite{Mochizuki_et_al__1999__aap__44TiitseffectivedecayrateinyoungsupernovaremnantsanditsabundanceinCassiopeiaA}.
Unfortunately, distribution and composition of dust in a SNR, though
very important for performing an analysis of the \Ti content in a SNR
as well as for the role of the SNR in the cosmic ray injection and
acceleration process, remain quite uncertain for Cas\,A and even more
so for other SNRs.

\section{Conclusions}
\label{Sek:Concl}

\cite{Chevalier_Oishi__2003__apjl__CassiopeiaAandItsClumpyPresupernovaWind}
suggested that Cas\,A underwent extensive mass loss, leading to a mass
of only $3-4\, \msol$ at the time of core collapse rather than the
original zero-age main-sequence mass of $20-25\, \msol$.  The
relatively slowly moving stellar wind formed a dense environment.
\cite{Woosley_et_al__1993__apj__Theevolutionofmassivestarsincludingmassloss-Presupernovamodelsandexplosion,Young_et_al__2006__apj__ConstraintsontheProgenitorofCassiopeiaA,Claeys_et_al__2011__aap__BinaryprogenitormodelsoftypeIIbsupernovae}
attribute these conditions to a binary companion aiding the mass loss
rather than the radiatively driven wind from a $20-25\, \msol$
progenitor.  This idea is also supported by the recent study of
\cite{Ouchi_Maeda__2017__apj__RadiiandMass-lossRatesofTypeIIbSupernovaProgenitors}
of the binary scenario for SN\,IIb. Such a scenario, if true, leads to
the existence of SN progenitors inside dense circumstellar matter (CSM
or CDS) produced by the extensive pre-supernova mass loss ($\dot{M}
>10^{-4}\, \msol\, \mathrm{yr}^{-1}$), a scenario that might be well
applied to Cas\,A.

The presence of large dust masses in and near SNRs might explain the
paucity of SNR-related \Ti detections in the Milky Way discussed by
\cite{The_et_al__2006__aap__Are44Ti-producingsupernovaeexceptional}.
Future observations of young galactic SNRs have to be carried out in
both energy bands of the \Ti lines at 68 and 78\,keV and at
1.157\,MeV, as well as at IR and mm wavelengths.  Only such
contemporaneous, multi-wavelength measurements can provide key
information on the dust content in and around SNRs.

We conclude by stating that the fluxes of \Ti decay lines measured at
68 and 78\,keV and 1,157\,MeV in time sequence measurements, together
with contemporaneous X-Ray, optical, IR, and mm measurements are very
important to shed light on the time sequence of dust formation during
the evolution of a SN to a SNR, to constrain the relative distribution
of dusty Fe and Ti in SNRs, and to verify the SN explosion mechanism
and the SNR asymmetry origin.

\section{Acknowledgements}
\label{Sek:Ackno}

MO acknowledges support from the European Research Council (grants
CAMAP-259276 and EUROPIUM-667912), from the Deutsche
Forschungsgemeinschaft through Sonderforschungsbereich SFB 1245
"Nuclei: From fundamental interactions to structure and stars", and
from the Spanish Ministry of Economy and Competitivity and the
Valencian Community under grants AYA2015-66899-C2-1-P and
PROMETEOII/2014-069, respectively.


\end{document}